\documentclass[runningheads]{llncs}
\usepackage{graphicx}
\usepackage{subcaption}
\usepackage{xcolor}
\usepackage{cite}
\usepackage{hyperref}
\hypersetup{
    colorlinks=true,
    linkcolor=blue,
    urlcolor=blue,
    citecolor=blue,
    pdfborder={0 0 0} 
}
\usepackage{comment}
\usepackage{amsmath, amsfonts}

\begin{document}
\title{Joint Multi-Echo/Respiratory Motion-Resolved Compressed Sensing Reconstruction of Free-Breathing Non-Cartesian Abdominal MRI}
\titlerunning{Joint Multi-Echo/Respiratory Motion-Resolved MRI Reconstruction}
%
\author{Youngwook Kee$^{1}$, MungSoo Kang$^{1}$, Seongho Jeong$^{1,2}$, and Gerald Behr$^{1}$}
\authorrunning{Y. Kee et al.}
%
\institute{$^1$Memorial Sloan Kettering Cancer Center, New York, NY, USA\\
$^2$Kyungpook National University, South Korea \\ \email{\{keey, kangm2, jeongs1, behrg\}@mskcc.org; setdotpy@knu.ac.kr}}
\maketitle              
\begin{abstract} 
We propose a novel respiratory motion-resolved MR image reconstruction method that jointly treats multi-echo k-space raw data. Continuously acquired non-Cartesian multi-echo/multi-coil k-space data with free breathing are sorted/binned into the motion states from end-expiratory to end-inspiratory phases based on a respiratory motion signal. Temporal total variation applied to the motion state dimension of each echo is then coupled in the $\ell_2$ sense for joint reconstruction of the multiple echoes. Reconstructed source images of the proposed method are compared with conventional echo-by-echo motion-resolved reconstruction, and R2* of the proposed and echo-by-echo methods are compared with respect to a clinical reference. We demonstrate that inconsistency between echoes is successfully suppressed in the proposed joint reconstruction method, producing high-quality source images and R2* measurements compared to clinical reference.

\end{abstract}
\section{Introduction}
\label{sec:introduction}

In multi-echo gradient-echo (mGRE) MRI, k-space raw data are acquired at multiple time points (echo times or TEs). Reconstructed images at different TEs are utilized in quantitative tissue parameter mapping, such as proton density fat fraction (PDFF)~\cite{dixon1984simple, reeder2005iterative}, R2* relaxation rate~\cite{hernando2013multipeak}, or quantitative susceptibility mapping (QSM)~\cite{wang2015quantitative, kee2017quantitative}. Clinical applications include fatty liver disease~\cite{li2018current} and hepatic iron overload assessment~\cite{hernando2014quantification}. Recent advances in non-Cartesian k-space sampling trajectories have facilitated free-breathing mGRE MRI \cite{kee2021free, armstrong2018free}, obviating the need for patients to hold their breath which is challenging for children and some adults. In addition, respiratory motion-resolved compressed sensing (CS) reconstruction, such as XD-GRASP~\cite{feng2016xd}, has demonstrated to enable motion-resolved PDFF/R2*/QSM~\cite{schneider2020free, kang2022ismrm, kang2022ismrm_2} mitigating the confounding factor of respiratory motion.

Despite these advances, applying respiratory motion-resolved CS reconstruction to highly-undersampled mGRE data is not straightforward. The question arises as to what mathematical machinery needs to be devised/incorporated for joint reconstruction of multiple echoes to prevent potential “echo inconsistency” produced by echo-by-echo CS reconstruction that necessarily requires iterative optimization. Little attention has been paid to the optimal treatment of highly-undersampled mGRE imaging data, which is crucial in motion-tolerant quantitative body MRI. In this paper, inspired by vectorial/collaborative total variation (TV) in color image denoising/deblurring/inpainting \cite{duran2016collaborative,goldluecke2012natural}, we propose a novel treatment for mGRE image reconstruction where all echoes are jointly reconstructed.

The remainder of this paper is organized as follows: Section \ref{sec:theory} describes the details of the proposed joint multi-echo/multi-coil/respiratory motion-resolved image reconstruction method with numerical implementation, Section \ref{sec:methods} describes the methods and materials used to demonstrate the effectiveness of the proposed method, Section \ref{sec:results} reports the results and compares them with existing methods, and finally, Section \ref{sec:discussion_conclusion} povides a discussion of our findings and a conclusion.

\section{Theory}
\label{sec:theory}

\subsection{Problem Formulation}

Let $N$ be the image size, $C$ be the number of coils, $E$ be the number of echoes, $T$ be the number of motion states, and $M$ be the number of measurements in k-space. We are interested in reconstructing $u_k \in \mathbb{C}^{N \times T}$ in image space from $y_{j,k} \in \mathbb{C}^{\lfloor M/T \rfloor \times T}$ in k-space for $k=1, \dots, E$ and $j=1, \dots, C$. Notice that the total number of k-space measurements $M$ is assumed to be sorted/binned into $\lfloor M/T \rfloor \times T$ from end-expiratory to end-inspiratory motion state according to a motion signal \cite{feng2016xd, kang2022ismrm, kang2022ismrm_2}. Therefore, each motion state is highly undersampled by a factor of $T$, and the following model-based MRI reconstruction is considered:
\begin{align} \label{MRI_eq1}
    \underset{u_1, \dots, u_E}{\text{minimize}} \,\frac{1}{2}\sum_{j=1}^{C}\sum_{k=1}^{E} ||\sqrt{D}(FS_j u_k - y_{j,k})||_F^2 + \lambda \cdot {\cal R}(u_1, \dots, u_E),
\end{align}
where $S_j: \mathbb{C}^{N \times T} \to \mathbb{C}^{N \times T}$ is the $j$-th coil sensitivity map, $F: \mathbb{C}^{N \times T} \to \mathbb{C}^{\lfloor M/T \rfloor \times T}$ is the nonuniform Fourier transform, and $D: \mathbb{C}^{\lfloor M/T \rfloor \times T} \to \mathbb{C}^{\lfloor M/T \rfloor \times T}$ is the density compensation factor \cite{ong2019accelerating, pruessmann2001advances}. $||\cdot||_F$ is the Frobenius norm in complex space and ${\cal R}(u_1, \dots, u_E)$ is the regularization term that can be defined via $\ell_2$ coupling along the echo as follows.
\begin{align} \label{MRI_eq2}
    {\cal R}(u_1, \dots, u_E) &= \sum_{{\bf x}} \left(|\partial_t u_1({\bf x})|^2 + |\partial_t u_2({\bf x})|^2 + \cdots + |\partial_t u_E({\bf x})|^2\right)^{1/2} \nonumber \\
    &= ||\partial_t {\bf u}||_{2,1},
\end{align}
where ${\bf u} = (u_1, \dots, u_E)$ and $\partial_t$ is the partial derivative or finite difference along the motion dimension (time). This is known as collaborative/vectorial TV \cite{duran2016collaborative, goldluecke2012natural}, initially considered in RGB color image processing. \emph{Here, we make a key observation that multiple echoes in mGRE data can be combined in the $\ell_2$ sense for joint multi-echo/respiratory motion-resolved CS reconstruction (henceforth, joint TE/MR recon). This approach takes into account the echo signal evolution, which is currently absent in the conventional echo-by-echo motion-resolved reconstruction (henceforth, echo-by-echo MR recon)}.

\subsection{Numerical Optimization}

The collaborative/vectorial TV with $\ell_2$ coupling is convex in $\bf u$, and we rewrite \eqref{MRI_eq1} with the dualization of the data consistency term \cite{chambolle2011first, kee2017primal} as follows.
\begin{align} \label{MRI_eq3}
    \min_{u_1, \dots, u_E} \max_{\substack{(\xi_1, \dots, \xi_E) \in K \\ \zeta_{1,1}, \dots, \zeta_{C,E}}}  \left( \sum_{j=1}^{C} \sum_{k=1}^E \langle \sqrt{D}(FS_ju_k - y_{j,k}), \zeta_{j,k} \rangle - \frac{1}{2}||\zeta_{j,k}||_2^2 \right)
    + \sum_{k=1}^E \langle \partial_t u_k, \xi_k \rangle,
\end{align}
where the convex set $K:=\{(\xi_1, \dots, \xi_E) \in \mathbb{C}^{N \times T} \times \cdots \times \mathbb{C}^{N \times T} : (|\xi_1|^2 + \cdots + |\xi_E|^2)^{1/2} \leq 1/\lambda\}$. The primal-dual hybrid gradient (PDHG) algorithm \cite{chambolle2011first} is suitable to solve \eqref{MRI_eq3}. By applying gradient ascent for the dual variables and gradient descent for the primal variable, we obtain the following update equations:
\begin{align}
    \xi_k^{n+1} &\gets \operatorname{proj}_{K}(\xi_k^n + \sigma \partial_t \bar{u}_k^n), \label{PDHG_1}\\
    \zeta_{j,k}^{n+1} &\gets \operatorname{prox}(\zeta_{j,k}^n + \sigma (\sqrt{D}(FS_j\bar{u}_k^n - y_{j,k}))), \label{PDHG_2}\\
    u_k^{n+1} &\gets u_k^n - \tau \left( \sum_{j=1}^C (\sqrt{D}FS_j)^H\zeta_{j,k}^n + \partial_t^H\xi_k^n \right), \label{PDHG_3}\\
    \bar{u}_k^{n+1} &\gets 2u_k^{n+1} - u_k^n, \label{PDHG_4}
\end{align}
for all $k$ and $j$. The projection onto $K$ and proximal operator are performed by
\begin{align}
    \xi_k \gets \frac{\tilde{\xi}_k}{\max(1, (|\xi_1|^2 + \cdots + |\xi_E|^2)^{1/2}/\lambda)}, \quad \zeta_{j,k} \gets \frac{\tilde{\zeta}_{j,k}}{1 + \sigma}. \label{PDHG_5}
\end{align}
The step sizes $\sigma$ and $\tau$ are both set to 1/8. The additional update step for the primal variable in \eqref{PDHG_4} is an extragradient step for convergence \cite{chambolle2011first}.

\paragraph{\underline{Remark I.}} The calculation of $(\sqrt{D}FS_j)^H\zeta_{j,k}^n$ for $j=1, \dots, C$, which is the most computationally intensive step, can be distributed across multiple GPUs using the Message Passing Interface (MPI) to achieve speedup.

\paragraph{\underline{Remark II.}} The $\ell_1$ coupling between echoes for ${\cal R}(u_1, \dots, u_E)$ can be used instead, which is expressed as
\begin{align} \label{MRI_eq3}
    ||\partial_t {\bf u}||_{1,1} = \sum_{{\bf x}} |\partial_t u_1({\bf x})| + |\partial_t u_2({\bf x})| + \cdots + |\partial_t u_E({\bf x})|.
\end{align}
Notice that there is \emph{no effective coupling} between the echoes as the gradient of the objective function with respect to each echo is independent from each other, possibly causing ``echo inconsistency'' in the final image. The only change in the PDHG algorithm is the convex set $K$ on which the dual variable $\zeta_{j,k}$ is projected, which is performed as $\xi_k \gets \tilde{\xi}_k / \max(1, |\xi_k|/\lambda)$.

\section{Methods}
\label{sec:methods}

\subsection{Synthetic Image (Toy Example)}
\label{subsec:syntehtic}

\begin{figure}[t!]
\includegraphics[width=\linewidth]{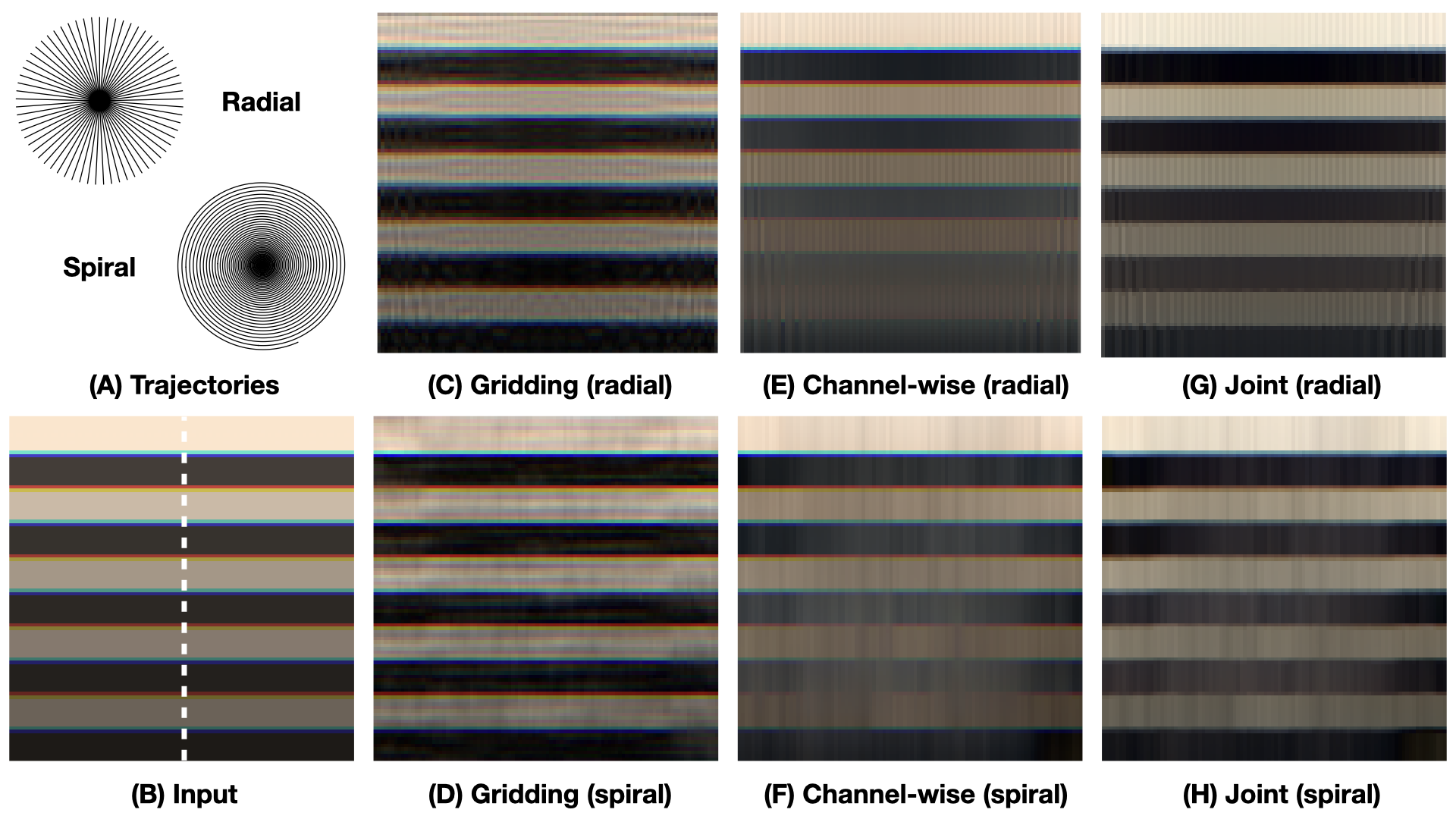}
\includegraphics[width=\linewidth]{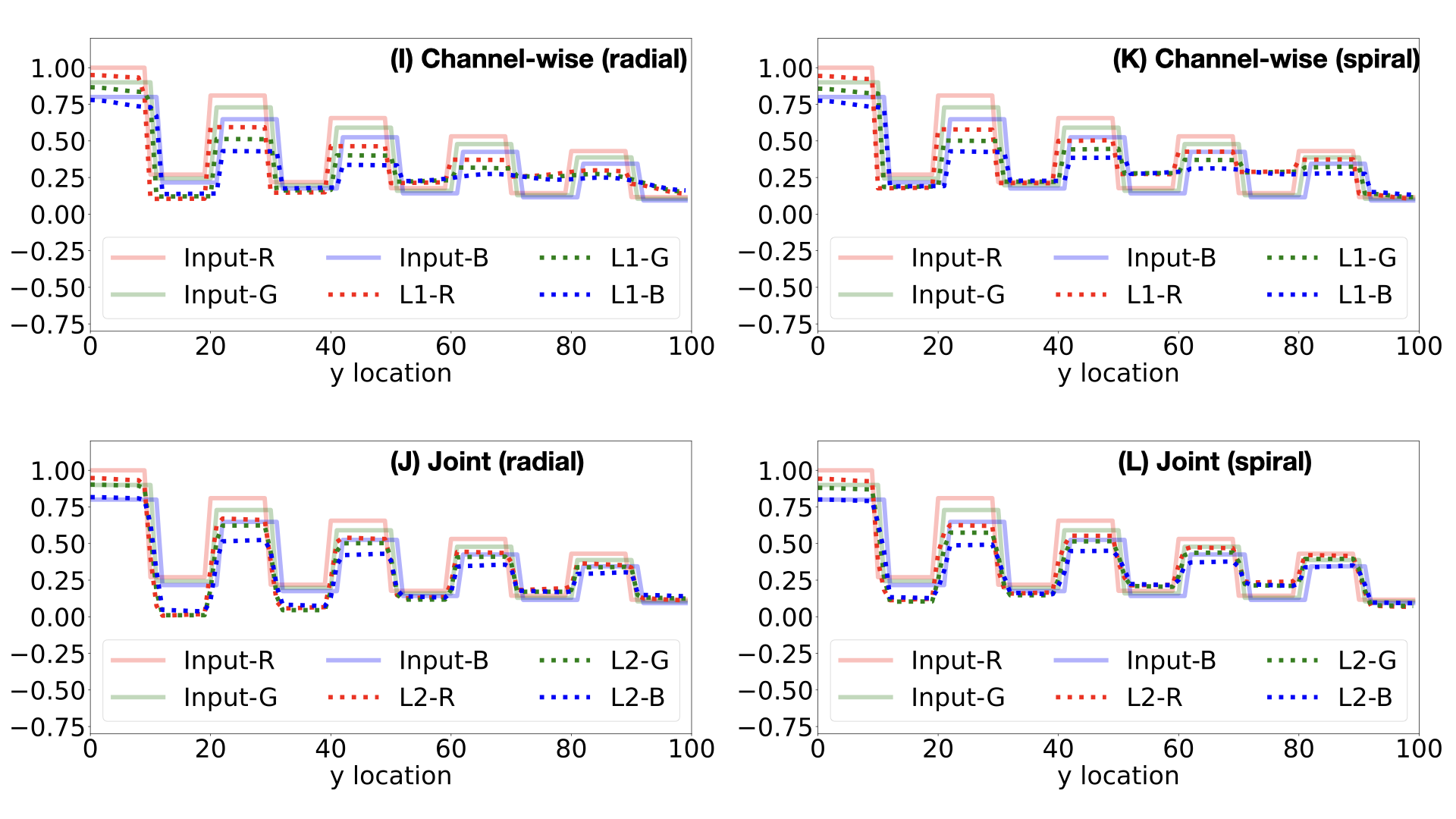}
\caption{Synthetic Image Experiment (see Section \ref{subsec:syntehtic} and \ref{sec:results} for description.).} \label{fig1}
\end{figure}

To illustrate the concept of collaborative/vectorial TV in the idea of joint TE/MR recon, we designed a synthetic image reconstruction experiment as follows: 2D radial and variable density spiral k-space sampling trajectories were designed using the following parameters: Gmax = 80 mT/m, SR = 200 T/m/s, FOV = 25 cm, in-plane resolution = 1$\times$1 mm$^2$, R = 6 (acceleration). The generated trajectories in Fig. \ref{fig1}(A) were used to sample the k-space (Fourier space) data of the input RBG image shown in Fig. \ref{fig1}(B) using the nonuniform Fourier transform (NUFFT). Notice that there is misalignment between the RGB color channels along the vertical axis, which was introduced by shifting the G and B channels. This is illustrated in Fig. \ref{fig1}(I-L), where transparent line profiles are drawn along the vertical dashed line in Fig. \ref{fig1}(B).

The R, G, and B channels can be considered as multi-echo images, where each channel represents the images at TE1, TE2, and TE3, respectively. The intensity profiles of each color channel, or equivalently each echo, can be viewed as representing different motion states. We formulate the following reconstruction problem: Let $m_j: (\Omega_{{\cal F}} \subset \mathbb{R}^2) \to \mathbb{C}$ be the $j$-th channel of the vector-valued function in Fourier space (k-space) $m: (\Omega_{{\cal F}} \subset \mathbb{R}^2) \to \mathbb{C}^3, (k_x, k_y) \mapsto (m_1(k_x,k_y), m_2(k_x,k_y),$ $m_3(k_x,k_y))$. The goal is to reconstruct ${\bf u}: (\Omega \subset \mathbb{R}^2) \to \mathbb{C}^3, (x, y) \mapsto (u_1(x,y), u_2(x,y), u_3(x,y))$ from the undersampled input Fourier samples $m$ by solving
\begin{align} \label{reconstruction_eq1}
    \underset{u_1, u_2, u_3}{\text{minimize}} \, \frac{1}{2} \sum_{k=1}^{3} \int_{\Omega_{\cal F}} \|\sqrt{d}({\cal F}u_k - m_k)\|_2^2 + \lambda \int_\Omega ||\partial_y {\bf u}||_1,
\end{align}
where ${\cal F}:\mathbb{C} \to \mathbb{C}$ is the nonuniform Fourier transform operator, and $d$ is the density compensation factor. The channels (or echoes) can be coupled with the $\ell_2$ norm as follows.
\begin{align} \label{reconstruction_eq2}
    \int_\Omega ||\partial_y {\bf u}||_2 = \int_\Omega \left( |\partial_y u_1|^2 + |\partial_y u_2|^2 + |\partial_y u_3|^2 \right)^{1/2}.
\end{align}
Note that the above variational formulation of generic image reconstruction problem \eqref{reconstruction_eq1} and \eqref{reconstruction_eq2} coincides with \eqref{MRI_eq1} and \eqref{MRI_eq2} for MRI reconstruction when $C=1$ (single coil). To evaluate the effectiveness of the proposed approach, we used PDHG to solve \eqref{reconstruction_eq1} and compared the results with gridding and channel-by-channel (or echo-by-echo) CS reconstructions.

\subsection{Human Subjects and Data Acquisition}

With IRB approval and informed consent, 7 subjects were recruited, including 4 adult patients with known and suspected iron overload and 3 healthy adult volunteers. The subjects underwent imaging on a 3T clinical MRI scanner (Signa Premier, GE Healthcare, Waukesha, WI) using a 3D multi-echo cones MRI method with free breathing (FB Cones) based on the implementation described in \cite{gurney2006design,kee2021free}. Subsequently, a commercially available Cartesian-based mGRE sequence (IDEAL-IQ, GE Healthcare, Waukesha, WI) was performed with a single breath-hold as a clinical reference (BH Cartesian).

One of the healthy subjects underwent two FB Cones acquisitions, one with regular breathing and the other with deep breathing. The imaging parameters for FB Cones were as follows: Initial TE/$\Delta$TE/TR = 0.032/1.4-1.5/11.4-11.5 ms; $\#$TEs = 6; FA = 3$^\circ$; resolution = 2$\times$2$\times$2 mm$^3$; rBW = 1106-1250 Hz/Px; readout duration = $\sim$1 ms. The imaging parameters for BH Cartesian were set to the clinical standard for iron quantification, which can be completed within a single breath-hold. These parameters include 4X parallel imaging acceleration along the phase and slice encoding directions and a slice thickness of 6 mm.

\begin{figure}[t!]
\centering
\includegraphics[width=\linewidth]{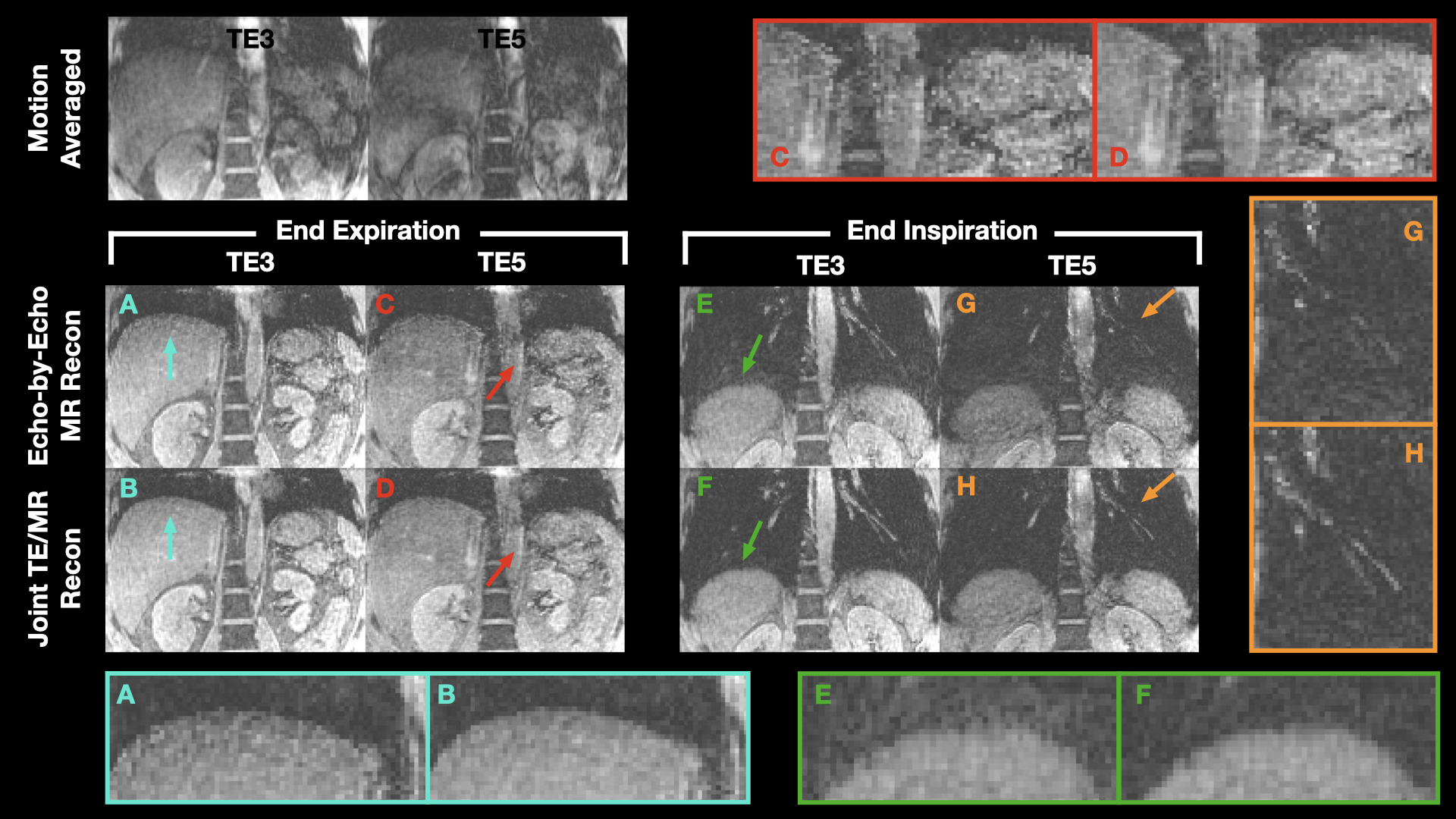}
\caption{Motion-averaged, echo-by-echo MR (A, C, E, and G) and join TE/MR (B, D, F, and H) reconstructions of end-expiration/inspiration of a healthy volunteer with deep breathing. The areas indicated by the color arrows are magnified in separate images and labeled with the same color bounding box.} \label{fig2}
\end{figure}

\subsection{Image Quality Assessment and ROI-based R2* Measurements}

To assess image quality, organ interfaces, such as the liver dome, pulmonary vasculature/bronchi, and liver/kidney interfaces, were closely examined visually and/or qualitatively. To facilitate this process, areas of interest were magnified, and the sharpness of the root sum of squares of the reconstructed multi-echo images (T2*w) was visually inspected using the `imgradient3' function in MATLAB. R2* maps were computed using a magnitude-based exponential fitting from reconstructed images from FB Cones as well as BH Cartesian. To quantify the R2* values in the liver parenchyma, we placed a 6-voxel radius region of interest (ROI) in the 3 consecutive coronal slices of the liver, avoiding large vessels that could affect the R2* measurements. Then, mean and std were calculated.

\section{Results}
\label{sec:results}

\begin{figure}[t!]
\centering
\includegraphics[width=\linewidth]{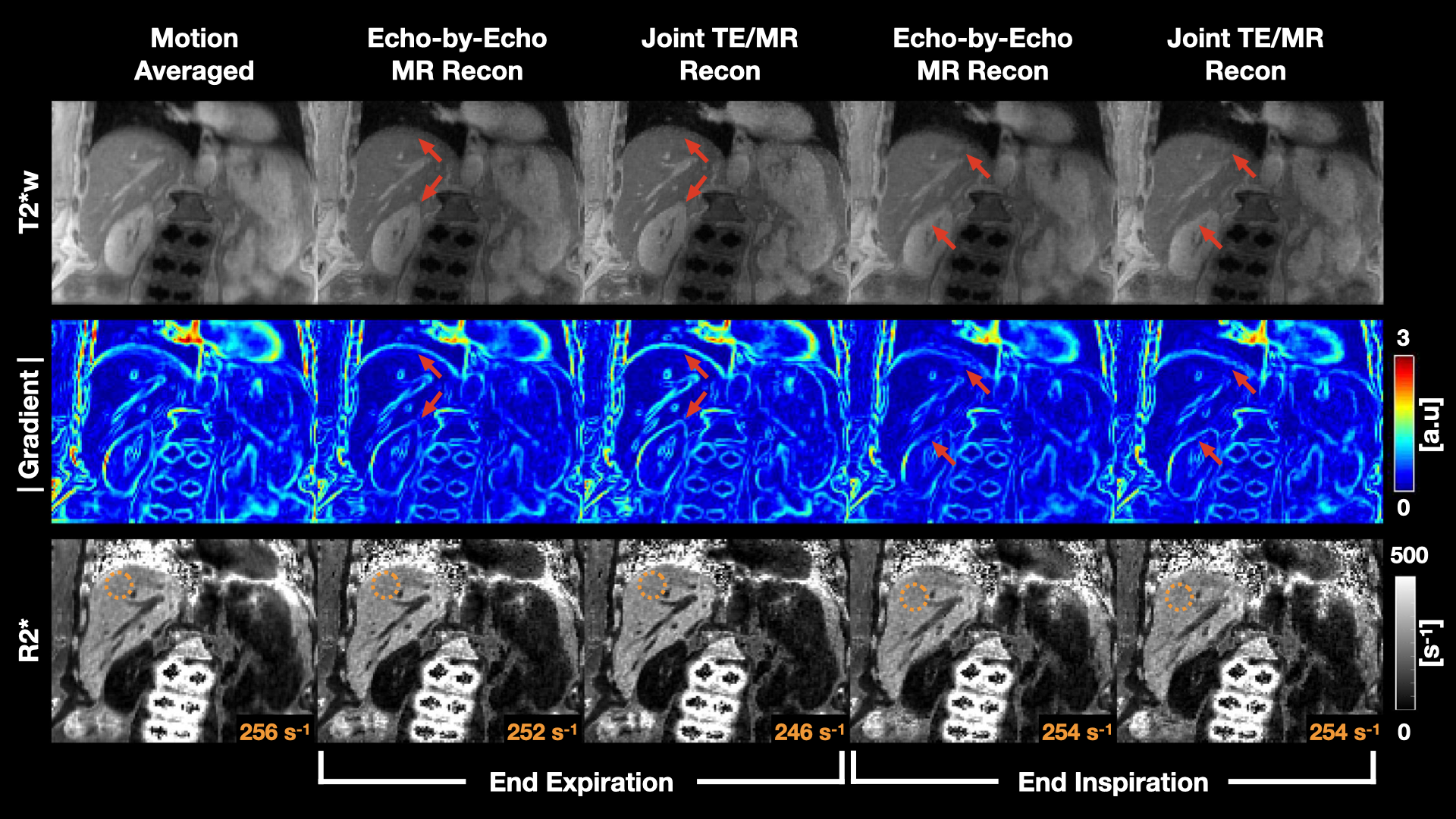}
\caption{T2*w, magnitude of gradients, and R2* maps for motion-averaged, echo-by-echo MR, and joint TE/MR reconstructions of a patient with iron overload are shown. The second and third columns show end-expiration, and the fourth and fifth columns show end-inspiration. ROI-based R2* measures are shown at the bottom of the R2* maps.} \label{fig3}
\end{figure}

Fig. \ref{fig1}(C-H) shows reconstructed RGB color images from the Fourier samples “acquired” with the radial and variable density spiral trajectories in Fig. \ref{fig1}(A). The color noise introduced by misalignment in the R, G, and B channels of the input image (Fig. \ref{fig1}(B)) was effectively suppressed in the joint reconstructions (Fig. \ref{fig1}(G, H)) regardless of the trajectory type. In contrast, gridding (Fig. \ref{fig1}(C, D)) and channel-wise reconstructions (Fig. \ref{fig1}(E, F)) preserved the color noise/misalignment. This difference is better appreciated in the line profiles (Fig. \ref{fig1}(I-L)) extracted from a vertical line passing through the center of the images, which is identical to the white dotted line in Fig. \ref{fig1}(B). The misaligned edges between the R, G, and B channels in the input image were reconstructed to have common edges in the joint reconstruction (Fig. \ref{fig1}(J, L)), as opposed to the channel-wise reconstruction (Fig. \ref{fig1}(I, K)).


Fig. \ref{fig2} shows reconstructed images of a healthy volunteer who performed deep breathing during the acquisition. Both echo-by-echo motion-resolved (echo-by-echo MR) and joint multi-echo/motion-resolved (joint TE/MR) methods demonstrated the capability to resolve end-expiratory/-inspiratory motion phases, in contrast to motion-averaged reconstruction. To facilitate a visual comparison between echo-by-echo MR and joint TE/MR reconstructions, areas of interest are indicated by colored arrows and magnified for closer examination. The joint TE/MR reconstruction provided a clearer visualization of the liver dome (Fig. \ref{fig2}(A, B, E, and F)) and pulomary vasculature/bronchi (Fig. \ref{fig2}(G, H)) when compared to the echo-by-echo reconstruction. In addition, the spurious noise present in the echo-by-echo MR reconstruction (Fig. \ref{fig2}(C)) was successfully suppressed in the joint TE/MR reconstruction (Fig. \ref{fig2}(D)).

Fig. \ref{fig3} displays T2*w, magnitude of gradients, and R2* maps of a patient with iron overload, and compares motion-averaged, echo-by-echo MR, and joint TE/MR reconstructions. The echo-by-echo MR and joint TE/MR reconstructions exhibit sharper image quality than the motion-averaged reconstruction, which shows a blurred liver dome. In the area of interest indicated by the red arrows, both end-expiratory and end-inspiratory phases of the joint TE/MR reconstruction exhibit visually sharper image quality than the echo-by-echo MR reconstruction. ROI-based R2* value obtained from the joint TE/MR reconstruction was closest to that obtained from BH Cartesian (230 s$^{-1}$; not shown).

The ROI-based R2* measurements of all the human subjects imaged for this study are reported in Table \ref{tab1}. As shown, the joint TE/MR reconstruction (second column from the right; show in bold) produced the lowest R2* values compared to the other reconstruction methods and was closest to those obtained from the clinical standard BH Cartesian method.

\begin{table}[t!]
\centering
\caption{ROI-based R2* measurements of healthy volunteers (HVs) and patients.}\label{tab1}

\vspace{0.1cm}

\begin{tabular}{|l|r|r|r|r|}
\hline
& FB Cones        & FB Cones     & FB Cones  & BH Cartesian \\
& motion-averaged & echo-by-echo MR & joint TE/MR &   reference\\
\hline
\hline
HV$\#$1           & 106.72 $\pm$ 19.05 & 70.18 $\pm$ 15.46 & \bf{65.51 $\pm$ 12.41} & 41.89 $\pm$ 14.81 \\
HV$\#$2           & 94.58 $\pm$ 21.98 & 76.19 $\pm$ 23.45 & \bf{71.73 $\pm$ 22.51} & 58.37 $\pm$ 17.34 \\
HV$\#$3 (regular) & 117.36 $\pm$ 25.89 & 87.59 $\pm$ 14.34 & \bf{80.79 $\pm$ 13.89} & 62.59 $\pm$ 18.93 \\
HV$\#$3 (deep)    & 246.08 $\pm$ 38.41 & 96.13 $\pm$ 25.53 & \bf{80.75 $\pm$ 17.68} & 39.35 $\pm$ 16.79 \\
Patient$\#$1      & 261.37 $\pm$ 28.54 & 258.94 $\pm$ 34.15 & \bf{252.25 $\pm$ 31.83} & 226.18 $\pm$ 32.07 \\
Patient$\#$2      & 311.98 $\pm$ 49.22 & 280.19 $\pm$ 49.81 & \bf{264.52 $\pm$ 47.50} & 224.97 $\pm$ 30.48 \\
Patient$\#$3      & 397.15 $\pm$ 51.95 & 393.47 $\pm$ 66.39 & \bf{378.08 $\pm$ 64.77} & 304.43 $\pm$ 58.74 \\
Patient$\#$4      & 489.12 $\pm$ 44.01 & 468.00 $\pm$ 52.00 & \bf{445.23 $\pm$ 56.12} & 400.32 $\pm$ 43.79 \\
\hline
\end{tabular}
\end{table}

\section{Conclusion}
\label{sec:discussion_conclusion}

In this paper, we have proposed a novel approach for handling multi-echo k-space raw data in respiratory motion-resolved CS reconstruction of highly-undersampled free-breathing 3D cones acquisition for liver R2* MRI. By making a key observation that multi-echo images in MRI are somewhat similar to the R, G, and B channels in a color image and demonstrating its effectiveness in a synthetic image reconstruction experiment, we have utilized vectorial/collaborative TV to suppress inconsistencies between echoes in \emph{in vivo} experiments. The proposed joint TE/MR reconstruction has been shown to successfully suppress “echo inconsistency” compared to echo-by-echo reconstructions, enabling 5D (3D space + 1D echo + 1D motion state) image reconstruction from continuously acquired non-Cartesian multi-echo k-space data. Future work will include exploring other convex/non-convex combinations of the echoes.

In conclusion, the proposed joint TE/MR reconstruction method successfully suppressed “echo inconsistency” observed in echo-by-echo motion-resolved CS reconstruction of free-breathing 3D multi-echo cones liver MRI.




%
%
%

%

\newpage
\subsection*{Suplementary Materials}

We provide the results of a phantom study performed as supplementary materials, which strongly supports the conclusion drawn from the \emph{in vivo} human subject study in the main manuscript.

\begin{figure}[h!]
\centering
\includegraphics[width=\linewidth]{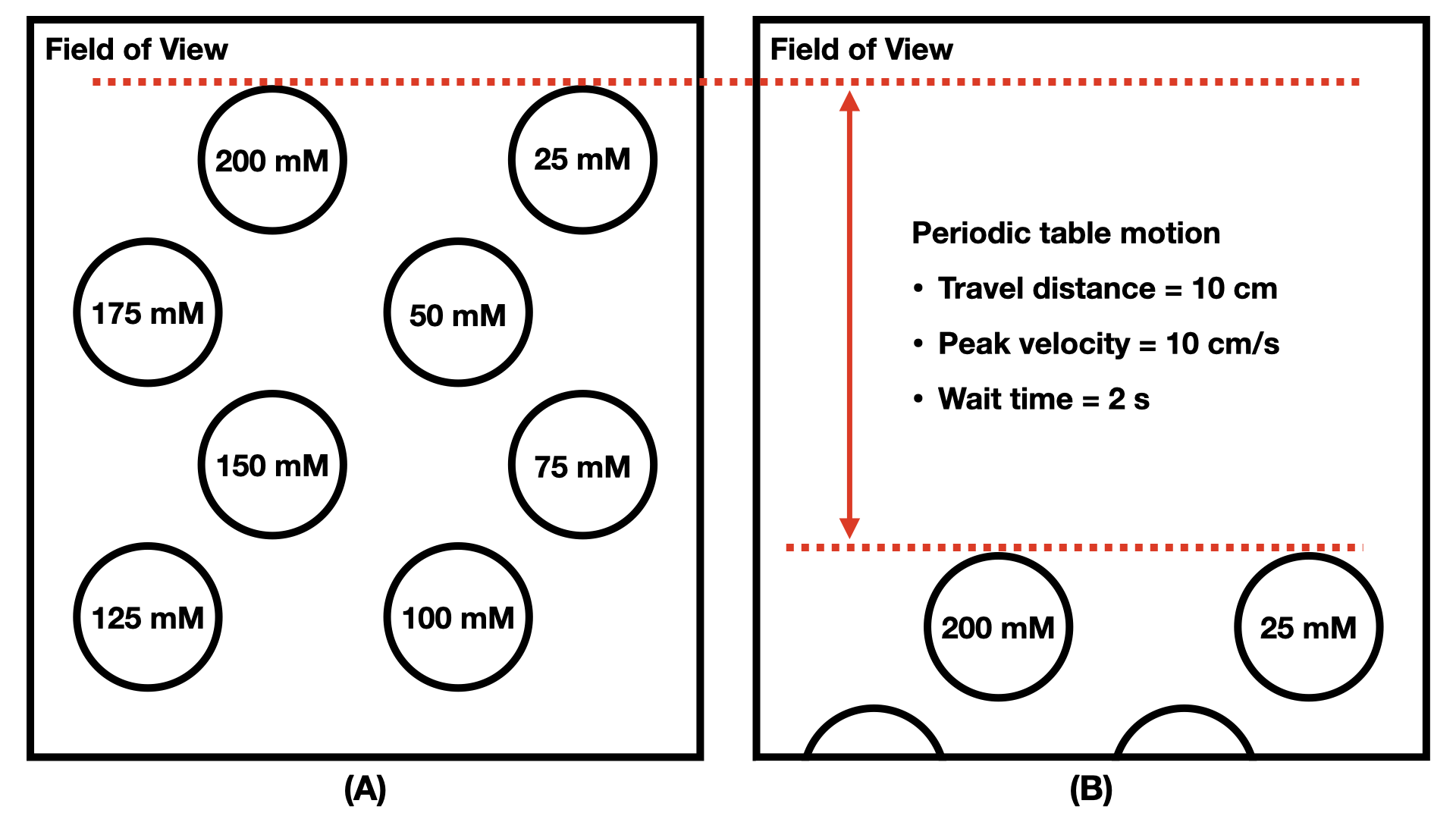}
\includegraphics[width=.98\linewidth]{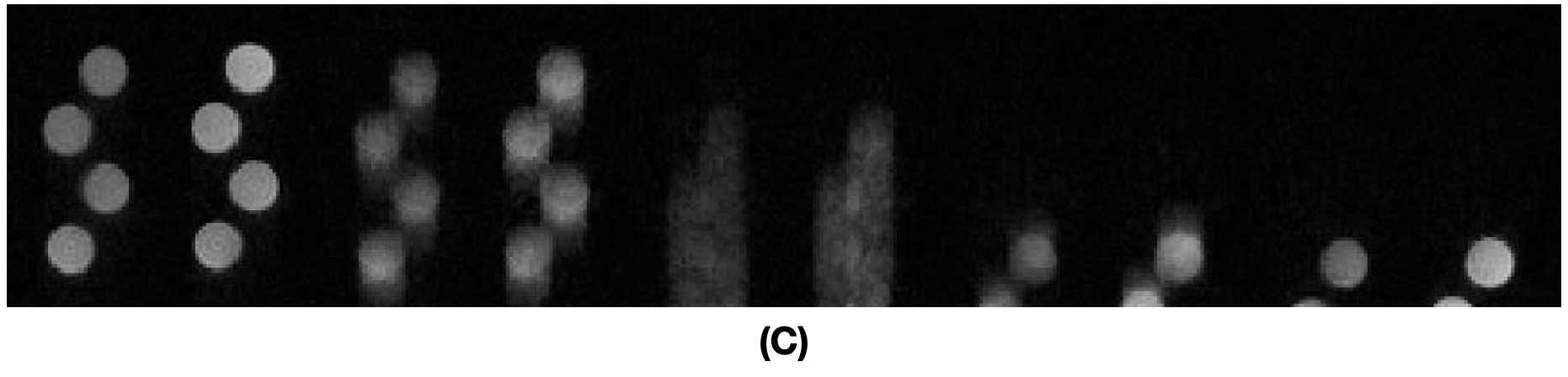}
\caption{Configuration of a phantom containing vials with gadolinium (Gd) solutions at different concentrations is illustrated in {\bf (A)}. The Gd solution (Gadavist, Bayer Pharmaceuticals, Wayne, NJ) was diluted with saline and 1$\%$ agarose gel solution to eight concentrations ranging from 25 to 200 mM. Each of the diluted Gd solutions (25, 50, 75, 100, 125, 150, 175, and 200 mM) was placed in a 50 mL Falcon tube and contained within a polystyrene foam container. During acquisition, periodic table motion was applied with the parameters prescribed in {\bf (B)}. Reconstructed images (at TE1) using the proposed method with five motion states are shown in {\bf (C)}. Note that the first column in {\bf (C)} may be considered the end-expiratory motion state in the \emph{in vivo} study reported in the main manuscript. In addition to the proposed method, echo-by-echo MR reconstruction and gridding reconstruction without table motion were also performed.}
\end{figure}

\begin{figure}[h!]
\centering
\includegraphics[width=\linewidth]{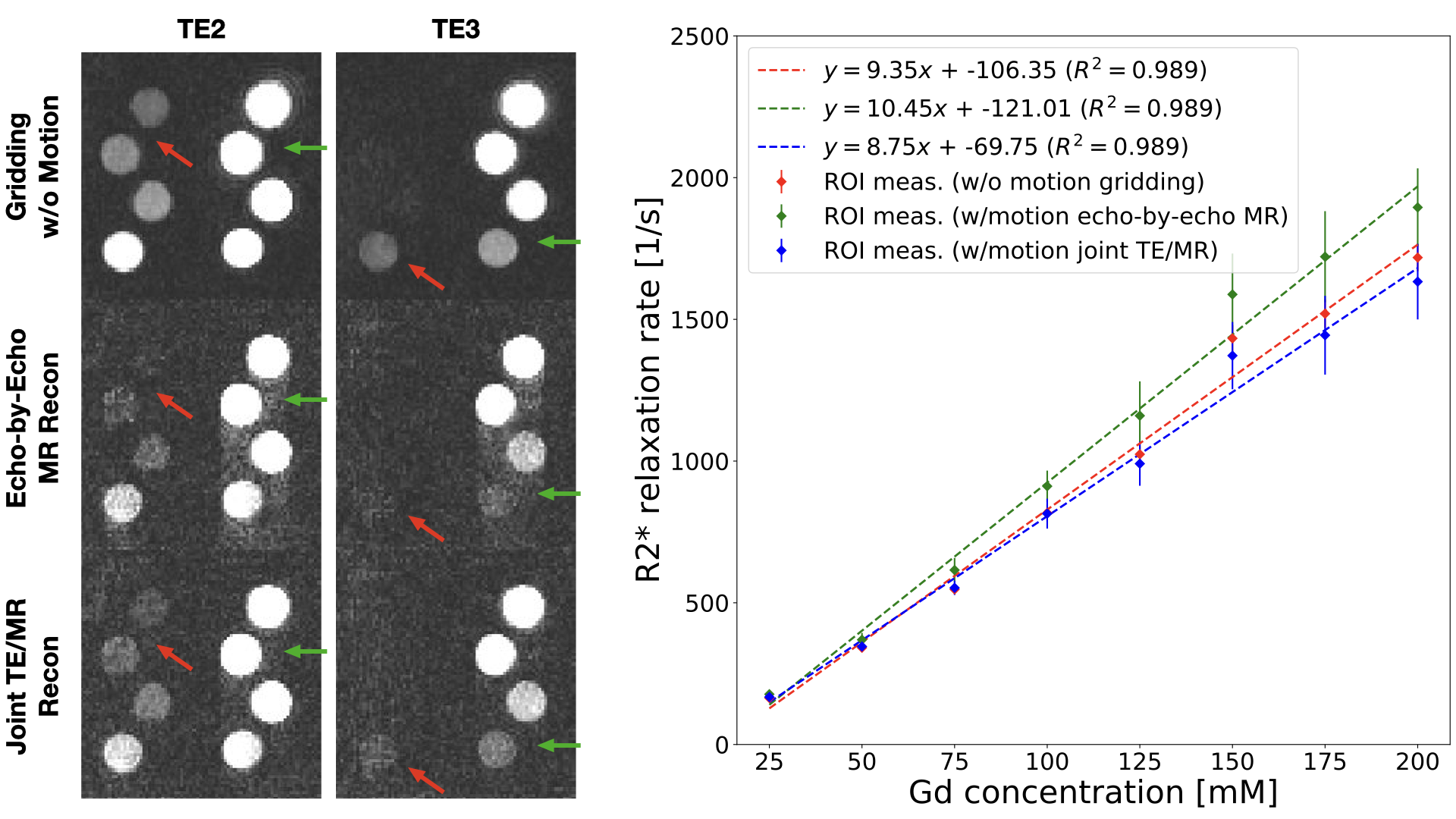}
\includegraphics[width=\linewidth]{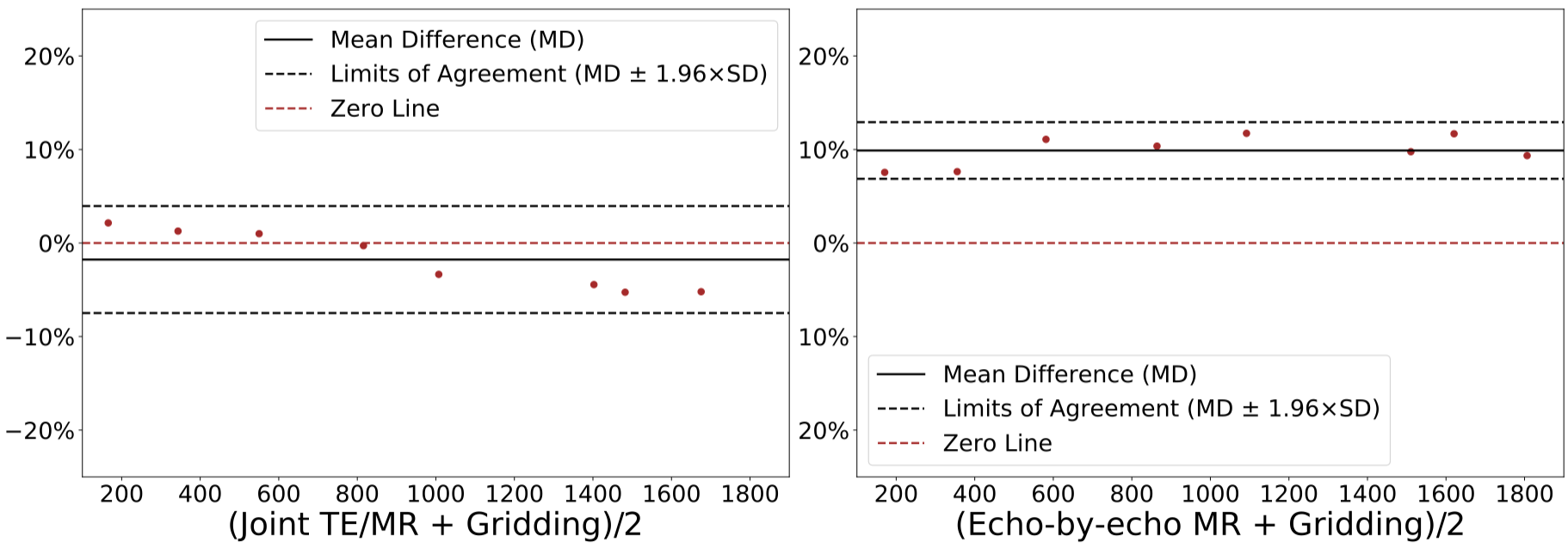}
\caption{Reconstructed images (gridding without table motion, echo-by-echo MR with table motion, and join TE/MR recons with table motion) at TE2 and TE3 are shown in the first and second columns of the top row. 
For the echo-by-echo MR and joint TE/MR reconstructions, the motion state where all the vials are inside the prescribed FOV, or that corresponds to end-expiratory motion state, is shown. 
As indicated by the red arrows, the echo-by-echo MR reconstruction exhibts poor SNR in the vials with high concentrations of Gd solution ([Gd]), compared to joint TE/MR reconstruction, which shows similar image quality to the gridding without motion. 
As also indicated by the green arrows, the residual \emph{unresolved} motion in the echo-by-echo MR reconstruction is clearly suppressed in the joint TE/MR reconstruction. 
These reconstruction artifacts manifest as inaccuracies in R2* measurements, as shown in the [Gd]-R2* scatter plot (third column of the top row) as well as in the Bland-Altman plots (bottom row). 
The ROI measurements from the joint TE/MR reconstruction are closer to those of the gridding reconstruction without bias than the echo-by-echo MR reconstruction.
Note that the vertical axis of the Bland-Altman plot is (joint TE/MR or echo-by-echo MR $-$ gridding)/gridding $\times$ 100 [$\%$].
}
\end{figure}
\end{document}